\documentclass[aps,showpacs,twocolumn]{revtex4}
\usepackage{epsfig}

\begin{document}

\title{Dynamical calculation of the $\Delta\Delta$ dibaryon candidates}

\author{Hongxia Huang$^{a}$, Jialun Ping$^a$, and Fan Wang$^b$}

\affiliation{$^a$Department of Physics, Nanjing Normal University,
Nanjing 210097, P.R. China}

\affiliation{$^b$Department of Physics, Nanjing University,
Nanjing 210093, P.R. China}

\begin{abstract}
We perform a dynamical calculation of the $\Delta\Delta$ dibaryon
candidates with $IJ^{P}=03^{+}$ and $IJ^{P}=30^{+}$ in the
framework of two constituent quark models: the quark delocalization
color screening model and the chiral quark model.
Our results show that the dibaryon resonances with $IJ^{P}=03^{+}$ and
$IJ^{P}=30^{+}$ can be formed in both models. The mass and width of
$IJ^{P}=03^{+}$ state are smaller than that of $IJ^{P}=30^{+}$ state due
to the one-gluon-exchange interaction between quarks. The resonance mass
and decay width of $IJ^{P}=03^{+}$ state in both models agree with that
of the recent observed resonance in the reaction $pn
\rightarrow d\pi^{0}\pi^{0}$. The $IJ^{P}=30^{+}$ $\Delta\Delta$ is
another dibaryon candidate with smaller binding energy and larger width.
The hidden-color channel coupling is added to the chiral
quark model, and we find it can lower the mass of the dibaryons by 10-20 MeV.
\end{abstract}


\maketitle

\setcounter{totalnumber}{5}

\section{\label{sec:introduction}Introduction}

The possibility of dibaryon states was first proposed by F. J.
Dyson and N. Xuong~\cite{Dyson} in 1964. However, this topic
got considerable attention only after R. Jaffe's prediction of
the $H$ particle in 1977~\cite{Jaffe}. All quark models, including
lattice QCD calculations, predict that in addition to
$q\bar{q}$ mesons and $q^{3}$ baryons, there should be
multiquark systems $(q\bar{q})^2$, $q^{4}\bar{q}$, $q^{6}$,
quark-gluon hybrids $q\bar{q}g$, $q^{3}g$, and
glueballs~\cite{Jaffe2}. A worldwide theoretical and
experimental effort to search for dibaryon states with and without
strangeness lasts for a long time. The $S=0$, $J^{P}=0^{-}$
$d^{\prime}$ dibaryon, which is hard to be explained by quark models~\cite{Ping},
was claimed by experiments in 1993 and disappeared years
later~\cite{dprime}. Our group showed that the $S=0$,
$I=0$, $J=3$ $d^{*}$ is a tightly bound six-quark system rather
than a loosely bound nucleus-like system of two
$\Delta$s~\cite{Goldman,QDCSM0,QDCSM1,QDCSM2}. An $S=-3$, $I=1/2$,
$J=2$ $N\Omega$ state was proposed as a high strangeness dibaryon
candidate~\cite{Goldman2}. Kopeliovich predicted high strangeness
dibaryons, such as the di-$\Omega$ with $S=-6$, using the flavor
$SU(3)$ Skyrmion model~\cite{Kopeliovich}. Zhang $et~al.$
suggested to search for the di-$\Omega$ in ultrarelativistic heavy
ion collisions~\cite{Zhang}. La France and Lomon predicted a
deuteron-like dibaryon resonance using $R$-matrix
theory~\cite{LaFrance} and measurements at Saclay seem to offer
experimental support for its existence~\cite{Lehar}. Despite
numerous claims, there has not been a well-established experimental
candidate for these dibaryon states.

However, the interest in the $H$ particle have been revived recently by
lattice QCD calculations of different collaborations,
NPLQCD~\cite{NPL} and HALQCD~\cite{HAL}. These two groups reported
that the $H$ particle is indeed a bound state at pion mass
larger than the physical one. Then, Carames and Valcarce examined
the $H$ particle within a chiral constituent quark model and
obtained a bound $H$ dibaryon with $B_{H}=7$ MeV~\cite{Carames}.

Recently, a pronounced resonance structure has been observed in
$pn$ collisions leading to two-pion production in the reaction $pn
\rightarrow d\pi^{0}\pi^{0}$, which suggests the presence of an
$IJ^{P} = 03^{+}$ subthreshold $\Delta\Delta$ resonance, called
henceforth $d^{*}$, with a resonance mass  $M = 2.37$ GeV and a
width $\Gamma \approx 70$ MeV~\cite{ABC1,ABC2}. The relatively
large binding energy of this state shows that it is much closer
to these interesting multiquark states than a loosely bound system
such as the deuteron. However, the width is remarkably smaller
than that given by a naive model estimate $\Gamma_{\Delta} \lesssim \Gamma
\lesssim 2\Gamma_{\Delta}$, where $\Gamma_{\Delta} \approx 120$
MeV.

According to Ref.~\cite{Dyson}, in addition to $d^{*}$, one should
also have a state with mirrored quantum numbers for spin and
isospin, $i.e.$ $IJ^{P} = 30^{+}$, called $D_{30}$ in Ref.~\cite{Dyson}.
Recently, M. Bashkanov $et~al.$ further pointed out that
the observation of the $d^{*}$ resonance state raises the
possibility of producing other novel six-quark dibaryon
configurations allowed by QCD and showed the $D_{30}$ state could
be regarded as manifestations of hidden-color six-quark
configurations in QCD~\cite{Bashkanov}. To what extent such kind
spin-isospin symmetry exists in hadron spectroscopy? It should be
an interesting check of the Goldstone boson exchange model where the
isospin triplet $\pi$ exchange interaction has the spin-isospin
symmetry~\cite{GL}. On the other hand, many former quark model calculations showed
that the mass of $IJ^{P} = 03^{+}$ $\Delta\Delta$ state was much smaller than that of
$IJ^{P} = 30^{+}$ $\Delta\Delta$ state because these models include the effective
gluon exchange. In the quark delocalization color screening model (QDCSM) the
$IJ^{P} = 03^{+}$ state is bound by 320 MeV, while the $IJ^{P} =
30^{+}$ state is bound by only 48 MeV~\cite{QDCSM2}; By using the
standard confinement and one gluon exchange (OGE) interaction
model, Maltman found the $IJ^{P} = 03^{+}$ state is bound by 260
MeV, while the $IJ^{P} = 30^{+}$ state is bound by only 30
MeV~\cite{Maltman}. Both results are in qualitative agreement with
the results of Oka and Yazaki~\cite{Oka1,Oka2},
Cvetic~\cite{Cvetic}, Valcarce~\cite{ChQM1} and Z. Y.
Zhang~\cite{Li}. This situation calls for
a more quantitative study of the $IJ^{P} = 30^{+}$ state.

Quantum chromodynamics (QCD) is widely accepted as the fundamental
theory of the strong interaction. However, the direct use of QCD
for low-energy hadronic interactions, for example, the
nucleon-nucleon ($NN$) interaction, is still difficult because of
the nonperturbative complications of QCD. QCD-inspired quark
models are still the main approach to study the baryon-baryon
interaction. The most common used quark model in the study of
baryon-baryon interaction is the chiral quark model
(ChQM)~\cite{ChQM1,ChQM2,ChQM3}, in which the $\sigma$ meson is
indispensable to provide the intermediate-range attraction.
Another quark-model approach is the quark delocalization color
screening model (QDCSM)~\cite{QDCSM0}, which has been developed
with the aim of understanding the well-known similarities between
nuclear and molecular forces despite the obvious energy and length
scale differences. In this model, the intermediate-range attraction
is achieved by the quark delocalization, which is like the
electron percolation in the molecules. The color screening is
needed to make the quark delocalization possible and it might be
an effective description of the hidden color channel coupling~\cite{Huang}.
Therefore to study the $D_{30}$ state with QDCSM is especially interesting
because its special relation to the hidden color channel effect.
We have showed both QDCSM and ChQM give a good
description of the $S$ and $D$ wave phase shifts of $NN$ $(IJ=01)$
scattering and the properties of deuteron~\cite{Chen_NN} despite
the difference of the mechanism of the $NN$ intermediate range attraction.
Recently, the $d^{*}$ resonance in $NN$ $D$-wave scattering were
re-studied with the QDCSM and ChQM~\cite{Ping_NN}. Both models
give an $IJ^{P} = 03^{+}$ $\Delta\Delta$ resonances reasonable well.
Therefore we will use these two models to calculate the mass and decay
width of the $D_{30}$ dibaryon, and compare the result with
the $d^{*}$ resonance, to check if there is a $D_{30}$ dibaryon
state. The hidden color channels are added to the ChQM to check
their effect in the $\Delta\Delta$ system.

The structure of this paper is as follows. A brief introduction of two
quark models is given in section II. Section III devotes to
the numerical results and discussions. The last section is a summary.

\section{Two quark models}
\subsection{Chiral quark  model}

The Salamanca version of ChQM is chosen as the representative of
the chiral quark models. It has been successfully applied to
hadron spectroscopy and $NN$ interaction. The model details can be
found in Ref.~\cite{ChQM1}. Only the Hamiltonian and parameters
are given here. The ChQM Hamiltonian in the nucleon-nucleon sector
is
\begin{widetext}
\begin{eqnarray}
H &=& \sum_{i=1}^6 \left(m_i+\frac{p_i^2}{2m_i}\right) -T_c
+\sum_{i<j} \left[
V^{G}(r_{ij})+V^{\pi}(r_{ij})+V^{\sigma}(r_{ij})+V^{C}(r_{ij})
\right],
 \nonumber \\
V^{G}(r_{ij})&=& \frac{1}{4}\alpha_s {\mathbf \lambda}_i \cdot
{\mathbf \lambda}_j
\left[\frac{1}{r_{ij}}-\frac{\pi}{m_q^2}\left(1+\frac{2}{3}
{\mathbf \sigma}_i\cdot {\mathbf\sigma}_j \right)
\delta(r_{ij})-\frac{3}{4m_q^2r^3_{ij}}S_{ij}\right]+V^{G,LS}_{ij},
\nonumber \\
V^{G,LS}_{ij} & = & -\frac{\alpha_s}{4}{\mathbf \lambda}_i
\cdot{\mathbf \lambda}_j
\frac{1}{8m_q^2}\frac{3}{r_{ij}^3}[{\mathbf r}_{ij} \times
({\mathbf p}_i-{\mathbf p}_j)] \cdot({\mathbf \sigma}_i+{\mathbf
\sigma}_j),
\nonumber \\
V^{\pi}(r_{ij})&=& \frac{1}{3}\alpha_{ch}
\frac{\Lambda^2}{\Lambda^2-m_{\pi}^2}m_\pi \left\{ \left[ Y(m_\pi
r_{ij})- \frac{\Lambda^3}{m_{\pi}^3}Y(\Lambda r_{ij}) \right]
{\mathbf \sigma}_i \cdot{\mathbf \sigma}_j \right.\nonumber \\
&& \left. +\left[ H(m_\pi r_{ij})-\frac{\Lambda^3}{m_\pi^3}
H(\Lambda r_{ij})\right] S_{ij} \right\} {\mathbf \tau}_i \cdot {\mathbf \tau}_j,  \\
V^{\sigma}(r_{ij})&=& -\alpha_{ch} \frac{4m_u^2}{m_\pi^2}
\frac{\Lambda^2}{\Lambda^2-m_{\sigma}^2}m_\sigma \left[ Y(m_\sigma
r_{ij})-\frac{\Lambda}{m_\sigma}Y(\Lambda r_{ij})
\right]+V^{\sigma,LS}_{ij}, ~~~~
 \alpha_{ch}= \frac{g^2_{ch}}{4\pi}\frac{m^2_{\pi}}{4m^2_u}
 \nonumber \\
V^{\sigma,LS}_{ij} & = & -\frac{\alpha_{ch}}{2m_{\pi}^2}
\frac{\Lambda^2}{\Lambda^2-m_{\sigma}^2}m^3_{\sigma} \left[
G(m_\sigma r_{ij})- \frac{\Lambda^3}{m_{\sigma}^3}G(\Lambda
r_{ij}) \right] [{\mathbf r}_{ij} \times ({\mathbf p}_i-{\mathbf
p}_j)] \cdot({\mathbf \sigma}_i+{\mathbf \sigma}_j),
\nonumber \\
V^{C}(r_{ij})&=& -a_c {\mathbf \lambda}_i \cdot {\mathbf
\lambda}_j (r^2_{ij}+V_0)+V^{C,LS}_{ij}, \nonumber
\\
V^{C,LS}_{ij} & = & -a_c {\mathbf \lambda}_i \cdot{\mathbf
\lambda}_j
\frac{1}{8m_q^2}\frac{1}{r_{ij}}\frac{dV^c}{dr_{ij}}[{\mathbf
r}_{ij} \times ({\mathbf p}_i-{\mathbf p}_j)] \cdot({\mathbf
\sigma}_i+{\mathbf \sigma}_j),~~~~~~ V^{c}=r^{2}_{ij},
\nonumber \\
S_{ij} & = &  \frac{{\mathbf (\sigma}_i \cdot {\mathbf r}_{ij})
({\mathbf \sigma}_j \cdot {\mathbf
r}_{ij})}{r_{ij}^2}-\frac{1}{3}~{\mathbf \sigma}_i \cdot {\mathbf
\sigma}_j. \nonumber
\end{eqnarray}
\end{widetext}
Where $S_{ij}$ is quark tensor operator, $Y(x)$, $H(x)$ and $G(x)$
are standard Yukawa functions, $T_c$ is the kinetic energy of the
center of mass, $\alpha_{ch} $ is the chiral coupling constant,
determined as usual from the $\pi$-nucleon coupling constant. All
other symbols have their usual meanings. The parameters of ChQM
are given in Table \ref{parameters}.

\subsection{Quark delocalization color screening model}
The model and its extension were discussed in detail in
Ref.\cite{QDCSM0,QDCSM1}. Its Hamiltonian has the same form as
Eq.(1), but without $\sigma$ meson exchange and a phenomenological
color screening confinement potential is used,
\begin{eqnarray}
V^{C}(r_{ij})&=& -a_c {\mathbf \lambda}_i \cdot {\mathbf
\lambda}_j [f(r_{ij})+V_0]+V^{C,LS}_{ij}, \nonumber
\\
 f(r_{ij}) & = &  \left\{ \begin{array}{ll}
 r_{ij}^2 &
 \qquad \mbox{if }i,j\mbox{ occur in the same } \\
 & \qquad \mbox{baryon orbit}, \\
 \frac{1 - e^{-\mu r_{ij}^2} }{\mu} & \qquad
 \mbox{if }i,j\mbox{ occur in different} \\
 & \qquad \mbox{baryon orbits}.
 \end{array} \right.
\end{eqnarray}
Here, $\mu$ is the color screening constant to be determined by
fitting the deuteron mass in this model. The quark delocalization
in QDCSM is realized by allowing the single particle orbital wave
function of QDCSM as a linear combination of left and right
Gaussian, the single particle orbital wave functions in the
ordinary quark cluster model,
\begin{eqnarray}
\psi_{\alpha}(\vec{S}_i ,\epsilon) & = & \left(
\phi_{\alpha}(\vec{S}_i)
+ \epsilon \phi_{\alpha}(-\vec{S}_i)\right) /N(\epsilon), \nonumber \\
\psi_{\beta}(-\vec{S}_i ,\epsilon) & = &
\left(\phi_{\beta}(-\vec{S}_i)
+ \epsilon \phi_{\beta}(\vec{S}_i)\right) /N(\epsilon), \nonumber \\
N(\epsilon) & = & \sqrt{1+\epsilon^2+2\epsilon e^{-S_i^2/4b^2}}. \label{1q} \\
\phi_{\alpha}(\vec{S}_i) & = & \left( \frac{1}{\pi b^2}
\right)^{3/4}
   e^{-\frac{1}{2b^2} (\vec{r}_{\alpha} - \vec{S}_i/2)^2} \nonumber \\
\phi_{\beta}(-\vec{S}_i) & = & \left( \frac{1}{\pi b^2}
\right)^{3/4}
   e^{-\frac{1}{2b^2} (\vec{r}_{\beta} + \vec{S}_i/2)^2}. \nonumber
\end{eqnarray}
The mixing parameter $\epsilon(S)$ is not an adjusted one but
determined variationally by the dynamics of the multi-quark system
itself. This assumption allows the multi-quark system to choose
its favorable configuration in the interacting process. It has
been used to explain the cross-over transition between hadron
phase and quark-gluon plasma phase~\cite{liu}. The model
parameters are fixed as follows: The $u,d$-quark mass difference
is neglected and $m_u$=$m_d$ is assumed to be exactly $1/3$ of the
nucleon mass, namely $m_u$=$m_d$=$313$ MeV. The $\pi$ mass takes
the experimental value. The $\Lambda$ takes the same values as in
Ref.\cite{ChQM1}, namely $\Lambda$=4.2 fm$^{-1}$. The chiral
coupling constant $\alpha_{ch}$ is determined from the $\pi NN$
coupling constant as usual. The other parameters b, $a_c$, $V_0$,
and $\alpha_s$ are determined by fitting the nucleon and $\Delta$
masses and the stability of nucleon size.
All parameters used are listed in Table \ref{parameters}.
In order to compare the intermediate-range
attraction mechanism, the $\sigma$ meson exchange in ChQM and
quark delocalization and color screening in QDCSM, the same values
of parameters: $b,~\alpha_s,~\alpha_{ch},~m_u,~m_\pi,~\Lambda$ are
used for these two models. Thus, these two models have exactly the
same contributions from one-gluon-exchange and $\pi$ exchange. The
only difference of the two models is coming from the short and
intermediate-range part, $\sigma$ exchange for ChQM, quark
delocalization and color screening for QDCSM.

\begin{table}[ht]
\caption{Parameters of quark models}
\begin{tabular}{lcc}
\hline\hline
 & {\rm ChQM} & ~~~~{\rm QDCSM}     \\
\hline
$m_{u,d}({\rm MeV})$        &  313    &  ~~~~313     \\
$b ({\rm fm})$              &  0.518  &  ~~~~0.518   \\
$a_c({\rm MeV\,fm}^{-2})$   &  46.938 & ~~~~56.755   \\
$V_0({\rm fm}^{2})$         &  -1.297 & ~~~~-0.5279 \\
$\mu ({\rm fm}^{-2})$       &   --     &  ~~~~0.45        \\
$\alpha_s$                  &  0.485  &  ~~~~0.485  \\
$m_\pi({\rm MeV})$          &  138    &  ~~~~138      \\
$\alpha_{ch}$               &  0.027  & ~~~~0.027   \\
$m_\sigma ({\rm MeV})$      &  675    &    ~~~~--          \\
$\Lambda ({\rm fm}^{-1})$   &  4.2    &  ~~~~4.2     \\
\hline\hline
\end{tabular}
\label{parameters}
\end{table}

\section{The results and discussions}

The resonating group method (RGM), described in more detail in
Ref.~\cite{RGM}, is used to calculate the masses and decay widths
of two-baryon states with $IJ^{P}=03^{+}$ and $IJ^{P}=30^{+}$.
The channels involved are listed in Table~\ref{channels}. Here the
baryon symbol is used only to denote the isospin, the superscript
denotes the spin, $2S+1$, and the subscript
``8" denotes color-octet, so $^{2}\Delta_{8}$ means the
$I, S = 3/2, 1/2$ color-octet state.

\begin{widetext}
\begin{center}
\begin{table}[h]
\caption{The two-baryon channels for states with $IJ^{P}=03^{+}$ $30^{+}$.}
\begin{tabular}{l|ccccccccccc}
\hline \hline
  & 1 & 2 & 3 & 4 \\
 $IJ^{P}=03^{+}$ & $\Delta\Delta(^{7}S_{3})$ & $NN(^{3}D_{3})$  & $\Delta\Delta(^{3}D_{3})$  & $\Delta\Delta(^{7}D_{3})$  \\
         & 5 & 6 & 7  & 8  & 9  & 10  \\
         & $^{2}\Delta_{8}~^{2}\Delta_{8}(^{3}D_{3})$  & $^{4}N_{8}~^{4}N_{8}(^{3}D_{3})$ & $^{4}N_{8}~^{2}N_{8}(^{3}D_{3})$
         & $^{2}N_{8}~^{2}N_{8}(^{3}D_{3})$  & $^{4}N_{8}~^{4}N_{8}(^{7}S_{3})$  & $^{4}N_{8}~^{4}N_{8}(^{7}D_{3})$  \\  \hline
 $IJ^{P}=30^{+}$ & 1 & 2 & 3  \\
         & $\Delta\Delta(^{1}S_{0})$ & $\Delta\Delta(^{5}D_{0})$ & $^{2}\Delta_{8}~^{2}\Delta_{8}(^{1}S_{0})$  \\  \hline
  \hline
\end{tabular}
\label{channels}
\end{table}
\end{center}
\end{widetext}

Because an attractive potential is necessary for forming bound
state or resonance, we first calculate the effective potentials of
the $S-$wave $\Delta\Delta$ states. The effective potential
between two colorless clusters is defined as,
\begin{equation}
V(s)=E(s)-E(\infty),
\end{equation}
where $E(s)$ is the diagonal matrix element
of the Hamiltonian of the system in the generating coordinate. The
effective potentials of the $S-$wave $\Delta\Delta$ for
$IJ^{P}=03^{+}$ and $IJ^{P}=30^{+}$ cases within two quark models
are shown in Fig. 1(a) and (b). From Fig. 1, we can see that
the potentials are attractive for both $IJ^{P}=03^{+}$ and
$IJ^{P}=30^{+}$ $\Delta\Delta$ states, and the attraction of
the $IJ^{P}=03^{+}$ state is larger than that of $IJ^{P}=30^{+}$
state in two models. The difference of attraction between
$IJ^P=03^+$ and $IJ^P=30^+$ in QDCSM is larger than that in ChQM.

\begin{figure}[ht]
\epsfxsize=3.3in \epsfbox{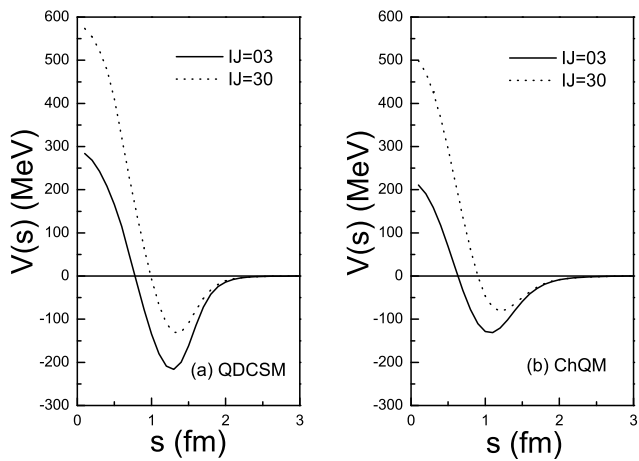}

\caption{The potentials of $S-$wave $\Delta\Delta$ for
$IJ^{P}=03^{+}$ and $IJ^{P}=30^{+}$ cases within two quark models.}
\end{figure}

In order to study what leads to the different effective
potentials between $IJ^{P}=03^{+}$ and $IJ^{P}=30^{+}$
$\Delta\Delta$ states, the contributions to the effective
potential from the kinetic energy, confinement,
one gluon exchange (OGE) and one boson exchange potentials
are calculated. We find that all the contributions are the same
between $IJ^{P}=03^{+}$ and $IJ^{P}=30^{+}$ $\Delta\Delta$ states,
except for the contribution from OGE potential, which are shown in
Fig. 2. From Fig. 2(a) and (b), we can see that OGE
potential of $IJ^{P}=03^{+}$ $\Delta\Delta$ state is attractive in
both QDCSM and ChQM, while OGE potential of $IJ^{P}=30^{+}$
$\Delta\Delta$ state is repulsive in both QDCSM and ChQM. Obviously,
the difference comes from the color-magnetic part of OGE
interaction ($V^{G}(r_{ij})$ in Eq.(1)). The color-magnetic
part contains the color and spin operator: $-{\mathbf
\lambda}_i \cdot {\mathbf \lambda}_j{\mathbf \sigma}_i\cdot
{\mathbf\sigma}_j$. The matrix elements of the operator for the
two states: $IJ^{P}=03^{+}$ and $IJ^{P}=30^{+}$, can be evaluated
as follows,
\begin{eqnarray}
V_{03} & = & -(6\sigma_{s}c_{s}+9\sigma_{s}c_{a}-6\sigma_{s}c_{s})  \\
V_{30} & = & -(6\sigma_{a}c_{s}+9\sigma_{s}c_{a}-6\sigma_{s}c_{s})
\end{eqnarray}
Here, $\sigma_{s}=1$, $\sigma_{a}=-3$, $c_{s}=\frac{4}{3}$,
$c_{a}=-\frac{8}{3}$. From Eq.(5) and Eq.(6), we can see that the
difference of the contributions from OGE
between $IJ^{P}=03^{+}$ and $IJ^{P}=30^{+}$ states comes from the
first term of these two expressions:
$-6\sigma_{s}c_{s}=-6\cdot1\cdot\frac{4}{3}=-8$ in $V_{03}$ and
$-6\sigma_{a}c_{s}=-6\cdot(-3)\cdot\frac{4}{3}=24$ in $V_{30}$,
which lead to the attractive OGE potential in $IJ^{P}=03^{+}$ case
and the repulsive OGE potential in $IJ^{P}=30^{+}$ case. So if one
do not include OGE interaction, the same result will be obtained
in $IJ^{P}=03^{+}$ and $IJ^{P}=30^{+}$ $S-$wave $\Delta\Delta$
states.

\begin{figure}[ht]
\epsfxsize=3.3in \epsfbox{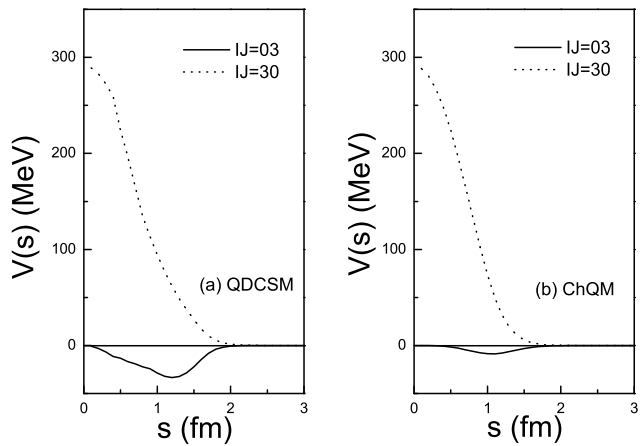}

\caption{The OGE potentials of $S-$wave $\Delta\Delta$ for
$IJ^{P}=03^{+}$ and $IJ^{P}=30^{+}$ cases within two quark models.}
\end{figure}

In order to see whether or not there is any bound state, a dynamic
calculation is needed. Here the RGM equation is employed.
Expanding the relative motion wavefunction between two clusters in
the RGM equation by gaussians, the integro-differential equation
of RGM can be reduced to algebraic equation, the generalized
eigen-equation. The energy of the system can be obtained by
solving the eigen-equation. In the calculation, the baryon-baryon
separation ($|\mathbf{s}_n|$) is taken to be less than 6 fm (to
keep the matrix dimension manageably small).

\begin{center}
\begin{table}
\caption{$\Delta\Delta$ or resonance mass $M$ and decay width
$\Gamma$, in MeV, in two quark models for the $IJ^{P}=03^{+}$
state.}
\begin{tabular}{c|cc|ccc}
\hline\hline
 \multicolumn{1}{c|}{}&\multicolumn{2}{|c|}{\rm QDCSM}&\multicolumn{3}{|c}{\rm ChQM}\\ \hline
     & ~~~~~{$sc.$}~~~ & ~~~{$4cc.$}~~~ &  ~~~~~{$sc.$}~~~ & ~~~{$4cc.$}~~~ & ~~~{$10cc.$}~~~  \\ \hline
   ~~~$M$ ~~~  &  2365   &   2357    &   2425   &  2413  &  2393  \\
   $\Gamma_{NN}$   &  --   &  14    &   --   &  14  &  14    \\
   $\Gamma_{inel}$   &  103   &  96    &   177   &  161  &  136    \\
   $\Gamma$   &  103   &  110    &   177   &  175  &  150    \\
   \hline\hline
\end{tabular}
\label{BE03}
\end{table}
\end{center}

For the $IJ^{P}=03^{+}$ state, the binding energy of
$\Delta\Delta$, resonance mass and decay width listed in Table
\ref{BE03} are taken from our previous calculation~\cite{Ping_NN}.
$sc.$ stands for the single channel $\Delta\Delta(^{7}S_{3})$ calculation;
$4cc.$ and $10cc.$ stand for channel-coupling calculations, ``4'' denotes
the four color-singlet channels listed in Table \ref{channels}, and ``10''
denotes the ten channels, four color-singlet channels and
six hidden-color channels listed in Table \ref{channels}.
$\Gamma_{NN}$ is the decay width of $\Delta\Delta(^{7}S_{3})
\rightarrow NN(^{3}D_{3})$; $\Gamma_{inel}$ is the inelastic
width caused by decaying $\Delta$s~\cite{Ping_NN} and $\Gamma$
stands for the total decay width
$\Gamma=\Gamma_{NN}+\Gamma_{inel}$. For the $IJ^{P}=30^{+}$ state,
since it cannot decay into $NN$ or $NN\pi$, but into the
$NN\pi\pi$ channel, we only calculate the inelastic width
$\Gamma_{inel}$ here. The binding energy of $IJ^{P}=30^{+}$
state and decay width $\Gamma=\Gamma_{inel}$ are listed
in Table \ref{BE30}. $sc.$ stands for the single channel
$\Delta\Delta(^{1}S_{0})$ calculation; channel coupling calculations
are denoted by $2cc.$ (two color-singlet channels) and $3cc.$ (two
color-singlet channels and one hidden-color channels).
There are several features which are discussed below.

\begin{center}
\begin{table}
\caption{$\Delta\Delta$ mass $M$ and decay width $\Gamma$, in MeV,
in two quark models for the $IJ^{P}=30^{+}$ state.}
\begin{tabular}{c|cc|ccc}
\hline\hline
 \multicolumn{1}{c|}{}&\multicolumn{2}{|c|}{\rm QDCSM}&\multicolumn{3}{|c}{\rm ChQM}\\ \hline
     & ~~~~~{$sc.$}~~~ & ~~~{$2cc.$}~~~ &  ~~~~~{$sc.$}~~~ & ~~~{$2cc.$}~~~ & ~~~{$3cc.$}~~~  \\ \hline
   ~~~$M$ ~~~  &  2430   &  2423     &   2457   &  2450  &  2440  \\
   $\Gamma$   &  185   &  175    &   228   &  216  &   200   \\
   \hline\hline
\end{tabular}
\label{BE30}
\end{table}
\end{center}

First, From Table \ref{BE03} and Table \ref{BE30}, we can see that
both the individual $IJ^{P}=03^{+}$ and $IJ^{P}=30^{+}$
$\Delta\Delta$ are bound in QDCSM and ChQM, which indicates that the
attraction between two $\Delta$s is strong enough to bind two
$\Delta$s together. However, the mass of $IJ^{P}=03^{+}$
state is smaller than that of $IJ^{P}=30^{+}$
state, due to the OGE interaction as mentioned above.
This result is in qualitative agreement with the results of our
previous study~\cite{QDCSM2}, Oka and Yazaki~\cite{Oka1,Oka2},
Cvetic~\cite{Cvetic}, Valcarce~\cite{ChQM1} and Z. Y.
Zhang~\cite{Li} as mentioned above. For the decay width, take the
QDCSM results as an example, the inelastic width $\Gamma_{inel}$
of $IJ^{P}=03^{+}$ state is 79 MeV smaller than that of
$IJ^{P}=30^{+}$ state, because of the smaller mass of
$IJ^{P}=03^{+}$ state. Although the $IJ^{P}=03^{+}$ state can decay to
$NN(^{3}D_{3})$ state, the decay width is only 14 MeV. So the total
decay width of the $IJ^{P}=03^{+}$ $\Delta\Delta$ is 110 MeV,
which is still 65 MeV smaller than that of the $IJ^{P}=30^{+}$
state. So the mass and width of the $IJ^{P}=03^{+}$ state
are both smaller than that of the $IJ^{P}=30^{+}$ state. The resonance
mass and decay width of the $IJ^{P}=03^{+}$ state indicate that
this resonance is a promising candidate for the observed
isoscalar ABC structure recently
reported by the CELSIUS-WASA Collaboration~\cite{ABC1} and
WASA-at-COSY Collaboration~\cite{ABC2}. The $IJ^{P}=30^{+}$
state is another possible six-quark dibaryon state and it might
be observed in proper experiments as discussed in Ref.~\cite{Bashkanov}.

Secondly, the similar results are obtained in ChQM. However, both
$IJ^{P}=03^{+}$ and $IJ^{P}=30^{+}$ states have smaller mass and
decay width in QDCSM than in ChQM. Our hidden color channel coupling calculation
in the $NN$ scattering shows that the color screening assumed in QDCSM is an
effective description of the hidden-color channel coupling effects~\cite{Huang}.
To check the effect of hidden-color channels coupling in ChQM, the hidden-color
channels are added to ChQM. For the $IJ^{P}=03^{+}$
state, the six hidden-color channels coupling lowers the ChQM resonance mass
by 20 MeV. For the $IJ^{P}=30^{+}$ state, the one hidden-color channel coupling
lowers the ChQM mass by 10 MeV. After including the hidden color channel coupling
the resonance masses in ChQM are closer to that in QDCSM.
So in the $\Delta\Delta$ system the hidden-color channel coupling effect is
also to increase the attraction, which is consistent with our previous conclusion
that the hidden-color channel coupling might be responsible
for the intermediate-range attraction of $NN$ interaction~\cite{Huang}.

\section{Summary}

In the present work, we perform a dynamical calculation of the
$\Delta\Delta$ dibaryon candidates with $IJ^{P}=03^{+}$ and
$IJ^{P}=30^{+}$ in the framework of QDCSM and ChQM. Our results
show that the attractions between two $\Delta$s is strong enough
to bind two $\Delta$s together for both $IJ^{P}=03^{+}$ and
$IJ^{P}=30^{+}$. However, the mass and width of the
$IJ^{P}=03^{+}$ state are smaller than that of the $IJ^{P}=30^{+}$
state due to the OGE interaction. The resonance mass and decay
width of the $IJ^{P}=03^{+}$ state indicate that this
$\Delta\Delta$ resonance is a promising candidate for the recent
observed one in the ABC effect. The $IJ^{P}=30^{+}$ $\Delta\Delta$ is
another possible six-quark dibaryon state and it might be observed in
proper experiments, such as $pp \rightarrow D_{30}\pi^{-}\pi^{-}
\rightarrow (pp\pi^{+}\pi^{+})\pi^{-}\pi^{-}$, which can be
done at COSY and JPARC~\cite{Bashkanov}.

The naive expectation of the spin-isospin symmetry is broken by the
effective one gluon exchange between quarks. The $d^*$ and $D_{30}$ states
searching will be another check of this gluon exchange mechanism and the
Goldstone boson exchange model.

QDCSM and ChQM obtained similar results. However, the
mass and decay width of $IJ^{P}=03^{+}$ and $IJ^{P}=30^{+}$
dibaryons in QDCSM are smaller than that in ChQM.
By including the hidden-color channels in ChQM, the resonance masses
are lowered by 10-20 MeV. This fact shows once more that the
quark delocalization and color screening used in QDCSM might be an effective
description of the hidden color channel coupling.

\section*{Acknowledgment}
This work is supported partly by the
National Science Foundation of China under Contract Nos. 11205091, 11035006 and 11175088.

\end{document}